\documentclass[sigconf,screen]{acmart}

\usepackage{graphicx} %
\usepackage{amsmath}
\usepackage{geometry}
\usepackage{array}
\usepackage{enumitem}
\usepackage[acronym]{glossaries}
\makeglossaries
\usepackage{multirow, makecell}
\usepackage{tikz}
\usepackage{adjustbox}
\usetikzlibrary{arrows.meta, positioning, shapes.geometric}

\newacronym{pgm}{PGM}{probabilistic graphical model}

\newacronym{nn}{NN}{neural network}
\newacronym{gbdt}{GDBT}{Gradient Boosted Decision Trees}
\newacronym{llm}{LLM}{large language model}
\newacronym{serp}{SERP}{search engine results page}
\newacronym{rctr}{RCTR}{rank-based CTR model}
\newacronym{dctr}{DCTR}{document-based CTR model}
\newacronym{rcm}{RCM}{random click model}
\newacronym{pbm}{PBM}{position-based model}
\newacronym{cm}{CM}{cascade model}
\newacronym{ubm}{UBM}{user browsing model}
\newacronym{dcm}{DCM}{dependent click model}
\newacronym{dbn}{DBN}{dynamic bayesian network model}
\newacronym{ncm}{NCM}{neural click model}
\newacronym{csm}{CSM}{click sequence model}
\newacronym{pscm}{PSCM}{partially sequential click model}
\newacronym{tcm}{TCM}{temporal click model}
\newacronym{thcm}{THCM}{temporal hidden click model}
\newacronym{emalgorithm}{EM}{expectation-maximization}
\newacronym{ips}{IPS}{inverse propensity scoring}
\newacronym{mle}{MLE}{maximum likelihood estimation}
\newacronym{ctr}{CTR}{click-through rate}
\newacronym{ccm1}{CCM}{carousel click model}
\newacronym{xpa}{XPA}{cross-positional attentions}
\newacronym{gubm}{GUBM}{grid-based user browsing model}
\newacronym{ccm2}{CCM}{click chain model}
\newacronym{cbcm}{CBCM}{comparison-based click model}
\newacronym{rbnn}{RBNN}{rank-biased neural network model}
\newacronym{drlc}{DRLC}{debiased reinforcement learning click model}
\newacronym{cacm}{CACM}{context-aware click model}
\newacronym{graphcm}{GraphCM}{graph-enhanced click model}
\newacronym{aicm}{AICM}{adversarial imitation click model}
\newacronym{fscm}{FSCM}{f-shape click model}
\newacronym{tacm}{TACM}{time-aware click model }
\newacronym{pctm}{PCTM}{probability click tracking model }
\newacronym{bbm}{BBM}{bayesian browsing model}
\newacronym{pcc}{PCC}{post-click click model}
\newacronym{gcm}{GCM}{general click model}
\newacronym{bss}{BSS}{bayesian sequential state model}
\newacronym{fetcm}{FE-TCM}{filter-enhanced transformer click model}

\newcommand{\headernodot}[1]{\vspace{1mm}\noindent\textbf{#1}}
\newcommand{\heading}[1]{\headernodot{#1.}}

\allowdisplaybreaks

\author{Jingwei Kang}
\orcid{0009-0003-9283-4060}
\affiliation{%
  \institution{University of Amsterdam}
  \city{Amsterdam}
  \country{The Netherlands}}
\email{j.kang@uva.nl}

\author{Maarten de Rijke}
\orcid{0000-0002-1086-0202}
\affiliation{%
  \institution{University of Amsterdam}
  \city{Amsterdam}
  \country{The Netherlands}}
\email{m.derijke@uva.nl}

\author{Santiago de Leon-Martinez}
\affiliation{%
  \institution{Brno University of Technology}
  \city{Brno}
  \country{Czechia}}
\additionalaffiliation{%
  \institution{Kempelen Institute of Intelligent Technologies}
  \city{Bratislava}
  \country{Slovakia}
}
\email{santiago.deleon@kinit.sk}
\orcid{0000-0002-2109-9420}

\author{Harrie Oosterhuis}
\orcid{0000-0002-0458-9233}
\affiliation{%
  \institution{Radboud University}
  \city{Nijmegen}
  \country{The Netherlands}}
\email{harrie.oosterhuis@ru.nl}

\copyrightyear{2025}
\acmYear{2025}
\setcopyright{cc}
\setcctype{by}
\acmConference[ICTIR '25]{Proceedings of the 2025 International ACM SIGIR Conference on Innovative Concepts and Theories in Information Retrieval (ICTIR)}{July 18, 2025}{Padua, Italy}
\acmBooktitle{Proceedings of the 2025 International ACM SIGIR Conference on Innovative Concepts and Theories in Information Retrieval (ICTIR) (ICTIR '25), July 18, 2025, Padua, Italy}
\acmDOI{10.1145/3731120.3744585}
\acmISBN{979-8-4007-1861-8/2025/07}


\title[Theory-Based Categorization and Design of Click Models]{Rethinking~Click~Models in Light of Carousel~Interfaces: Theory-Based Categorization and Design of Click Models}

\begin{document}

\begin{abstract}
      Click models are a well-established for modeling user interactions with web interfaces.
      Previous work has mainly focused on traditional single-list web search settings; this includes existing surveys that introduced categorizations based on the first generation of \gls{pgm} click models that have become standard.
      However, these categorizations have become outdated, as their conceptualizations are unable to meaningfully compare \gls{pgm} with \gls{nn} click models nor generalize to newer interfaces, such as carousel interfaces. %
      We argue that this outdated view fails to adequately explain the fundamentals of click model designs, thus hindering the development of novel click models. %

      This work reconsiders what should be the fundamental concepts in click model design, grounding them - unlike previous approaches - in their mathematical properties.
      We propose three fundamental key-design choices that explain what statistical patterns a click model can capture, and thus indirectly, what user behaviors they can capture.
      Based on these choices, we create a novel click model taxonomy that allows a meaningful comparison of all existing click models; this is the first taxonomy of single-list, grid and carousel click models that includes \glspl{pgm} and \glspl{nn}.
      Finally, we show how our conceptualization provides a foundation for future click model design by an example derivation of a novel design for carousel interfaces.
\end{abstract}

\maketitle
\glsresetall

\section{Introduction}
\heading{Click models}
Click models are probabilistic frameworks that are used to interpret user interactions, helping to infer relevance signals from real clicks \cite{10.1145/3018661.3018699,10.1145/3209978.3209986, 10.1145/3159652.3159732, 10.1145/2911451.2911537, 10.1145/3437963.3441794, 10.1145/1390334.1390392, 10.1145/1526709.1526711,chuklin_markov_rijke_click_models_2015}. 
During the past decade, numerous click models have been developed, primarily using \glspl{pgm} \cite{10.1145/1242572.1242643,10.1145/1341531.1341545,10.1145/1390334.1390392,10.1145/1498759.1498818,10.1145/1526709.1526711,10.1145/2124295.2124334,10.1145/2766462.2767712} or \glspl{nn} \cite{10.1145/2872427.2883033,10.1145/3209978.3210004}.
However, like most research in information retrieval, previous studies on click models have mostly focused on traditional web search with single-list layouts, implicitly assuming user interactions are limited to a single list and ignoring the context and more complex layout of the system the user is currently using.

\begin{figure}[t]
    \centering
    \includegraphics[width=\linewidth, alt={Netflix carousel interface}]{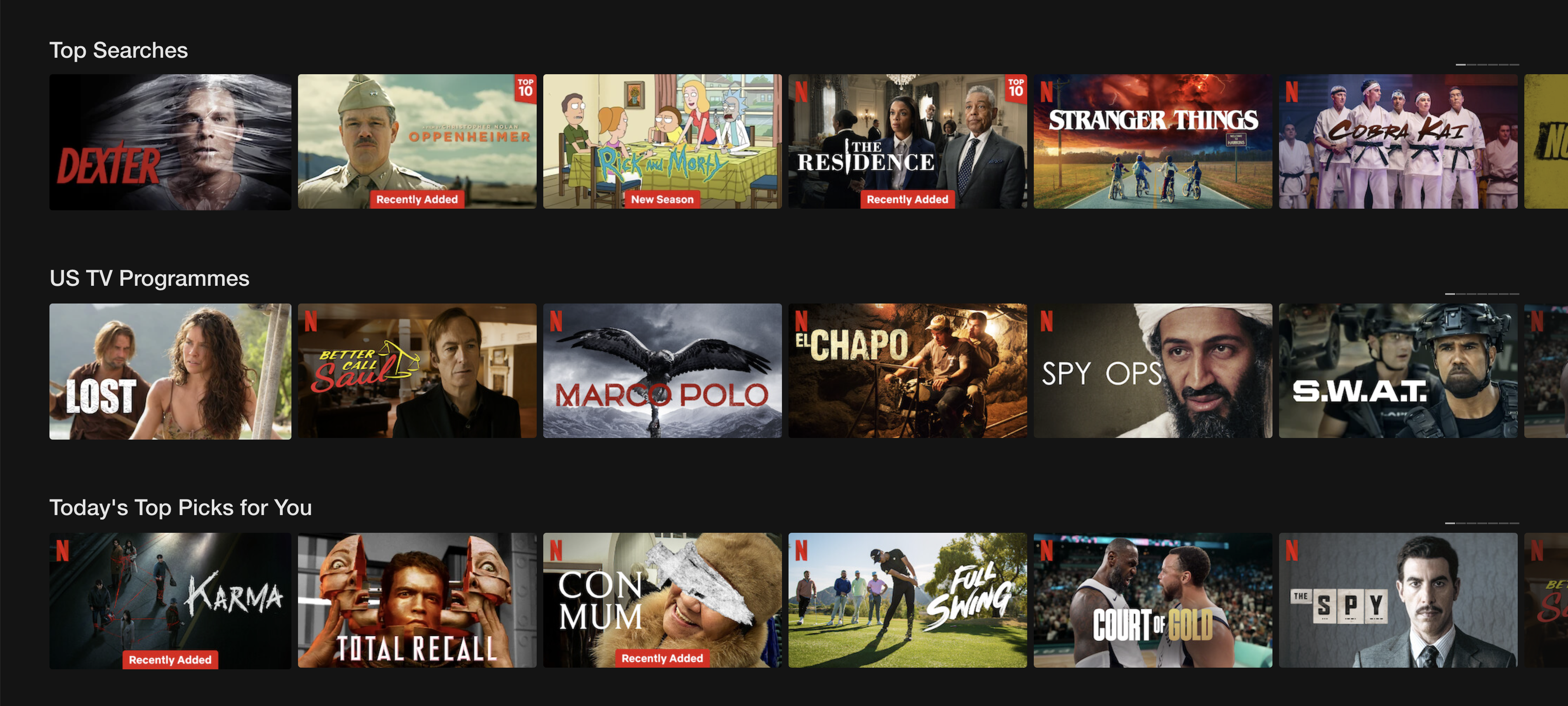}
  \vspace{-2\baselineskip}
    \caption{Example of media recommendations by Netflix presented in a carousel interface.}
    \label{fig:carousel}
  \vspace{-\baselineskip}
\end{figure}

\heading{Carousel interfaces}
In recent years, with the rise of modern streaming media services such as Netflix and Spotify, the way recommendations are presented has changed significantly \cite{10.1145/3383313.3412217, 10.1145/2959100.2959174, 10.1145/3289600.3291027}. 
Recommendations are no longer only presented as single vertical lists, but instead multiple horizontally scrollable (also known as \textit{swipeable}) lists are simultaneously displayed to the user in a vertical arrangement
, e.g., as shown in Figure~\ref{fig:carousel}
. 
In these so-called \emph{carousel} interfaces \cite{10.1145/3452918.3465493, 10.1145/3640544.3645223, 10.1145/3604915.3610638}, each list has a title (or topic) that represents its content, such as a specific type of movie or music, or a personalized category like \emph{Made for You} or \emph{Recently Played}.
Carousel interfaces have become increasingly popular for several reasons, in particular, their ability to support diverse user needs by presenting multiple collections of recommendations. The few existing user studies into carousel interfaces have shown that carousels may increase perceived diversity and novelty of items when compared to single lists and grids \cite{jannach_exploring_2021, starke_serving_2021,starke_examining_2023} and choice satisfaction \cite{starke_serving_2021}.
Furthermore, simulation experiments have shown that carousels are more efficient than ranked lists when scanning for items \cite{10.1145/3643709}.

\heading{Grid interfaces}
Another search and recommendation interface that is different from the traditional single-list interface is the grid interface, where items are displayed along the columns and rows of a grid layout~\cite{10.1145/1743666.1743736, doi:10.1080/10447318.2013.846790, 10.1145/2959100.2959150}.
The salient difference with carousel interfaces is that the rows in grid interfaces are not grouped or titled.
Accordingly, grid interfaces are most commonly used to display sets of items without further grouping, e.g., images in image-search or videos in social media video-recommendation.

\heading{Challenges}
The main existing survey on click models by \citet{chuklin_markov_rijke_click_models_2015} presents a conceptual framework for click models based on \glspl{pgm}, where different click model categories rely on distinct assumptions about user behaviors by pre-defining latent variables and hand-crafted dependencies.
However, this framework cannot naturally adapt to \gls{nn}-based click models, which use multi-layered \glspl{nn} to model user behavior instead of hand-crafted graphical network structures.
As a result, \gls{nn} click models do not have a place in the framework of \citet{chuklin_markov_rijke_click_models_2015}, later work that categorizes both \gls{pgm} and \gls{nn} click models place them in completely separate taxonomies~\cite{Liu2024}.
However, there are many similarities and differences between \gls{pgm} and \gls{nn} click models that go beyond the machine learning techniques they apply, but it appears that the field currently lacks a conceptualization of click models that enables such comparisons.

The lack of adaptability of the current conceptualization becomes even more apparent in the context of carousel interfaces, where user behavior differs significantly from traditional single-list interfaces.
Despite the many advantages of carousel interfaces, understanding user behavior on carousel interfaces remains a difficult challenge.
The presence of multiple lists (a.k.a.\ carousels) along with the ability to swipe to see more greatly increases the number of possible browsing actions.
Users can switch between horizontal exploration (within a carousel) and vertical exploration (between carousels), which may be heavily influenced by topic preferences.
Additionally, the visual content presented by the carousel interface is much richer than the \emph{ten blue links} of traditional web search, further encouraging users to explore.
However, this increase in presented information, along with the greater number of possible interactions with carousel interfaces, makes it much more difficult to model the decision-making process of the user.
Despite a recent attempt to propose a carousel click model \cite{10.1145/3643709}, there is no precedent for systematic carousel click model design, and the design possibilities seem innumerable.

In response, our goal in this paper is to re-establish a theoretical foundation and conceptualization for click models that properly captures both \gls{pgm} and \gls{nn} click models for single-list, grid and carousel interfaces.
Our work is guided by the following three research questions:
\begin{itemize}[leftmargin=*]
\item \emph{What mathematical concepts and properties should be used as the fundamental basis of the categorization of click models?}
\vspace{0.2\baselineskip}
\item \emph{What taxonomy does our theory-based approach produce for single-list, grid and carousel click models?}
\vspace{0.2\baselineskip}
\item \emph{How can our new approach to categorization guide future carousel click model design?}
\end{itemize}

\heading{Contributions} 
We make the following contributions:

\paragraph{A novel position that argues the categorization of click models should mainly concern the relations between observed variables.} 
When categorizing click models, we propose that the primary focus should be on observed variables that influence clicks, rather than rely on assumptions about user behaviors or latent variables.
This contrasts with the existing categorization of click models~\citep{chuklin_markov_rijke_click_models_2015}, where \glspl{pgm} are characterized by their latent variables and the specific user behavior they aim to model.
See Section~\ref{sec:per1} for more details.

\paragraph{An identification of three key design choices regarding their mathematical properties.}
We find that the design of any click model requires decisions on three key design choices, explicitly or implicitly: \textbf{global dependencies}, \textbf{sequentiality} and \textbf{factorization}.
Following the above principle, we argue that click model designs should use these choices as their starting point.
See Section~\ref{sec:choice} for more details.

\paragraph{A generalized taxonomy of click models.} Finally, following three key design choices, we propose a taxonomy that categorizes any - existing or future - single-list, grid or carousel interface click models into non-overlapping categories.
Because our taxonomy relies solely on the relations between observed variables, and makes no further assumptions about user behavior and latent variables, it is guaranteed to be applicable to any click model, i.e., any \gls{pgm} or \gls{nn}-based model, or any other future model.
See Section~\ref{sec:tax_carousel} for more details.

\paragraph{An example of carousel click model design.} Building on the taxonomy proposed above, we then introduce a specific carousel click model following three design choices to serve as a starting point for further research in this area.

\medskip\noindent
We believe that these contributions build a theoretical foundation for the conceptualization, categorization and design of click models and inspire future research in the underexplored field of designing click models for carousel interfaces.

\section{Related Work}
\subsection{Click models for single-list layouts}
Click models have been extensively studied in single-list web search, where search results are displayed in a top-to-bottom format.
These models serve a dual purpose: estimating document relevance from real click data via counterfactual estimation \cite{10.1145/3018661.3018699,10.1145/3209978.3209986, 10.1145/3159652.3159732, 10.1145/2911451.2911537, 10.1145/3437963.3441794} or maximum likelihood estimation \cite{10.1145/1390334.1390392, 10.1145/1526709.1526711,chuklin_markov_rijke_click_models_2015}, and predicting click probabilities to further simulate user clicks on \gls{serp} \cite{chuklin_markov_rijke_click_models_2015}.
Traditional click models primarily rely on the \gls{pgm} framework, which assumes various hypotheses about user behavior over unobserved latent variables (such as examination and satisfaction) and requires manual setting of the dependencies among these variables. 
Based on this framework, some models assume a sequential examination hypothesis in which users examine results from top to bottom; prominent models that rely on this assumption include the \gls{cm}~\cite{10.1145/1341531.1341545}, \gls{ubm}~\cite{10.1145/1390334.1390392}, \gls{dcm}~\cite{10.1145/1498759.1498818}, and \gls{dbn}~\cite{10.1145/1526709.1526711}. 
Some models avoid this assumption to account for non-sequential examination hypothesis, such as revisiting previously examined results. Examples include the \gls{thcm}~\cite{10.1145/2124295.2124334} and \gls{pscm}~\cite{10.1145/2766462.2767712}.

Although \glspl{pgm} are the most prominent basis for existing click models, with recent advances in deep learning~\cite{lecun2015deep}, \gls{nn}-based click models have also emerged.
Unlike \gls{pgm}-based methods that apply hand-crafted models, these neural approaches learn variable dependencies from data, representing behavior by a sequence of non-linear transformations. 
Examples include the \gls{ncm} \cite{10.1145/2872427.2883033} and \gls{csm} \cite{10.1145/3209978.3210004}, both of which aim to reduce the need for explicit assumptions about user examination patterns.

\citet{Liu2024} propose a taxonomy of \gls{pgm}-based click models that categorizes them based on assumptions about user examination.
They divide (web search) click models into \emph{position-based models} that assume that users examine the \gls{serp} sequentially from top to bottom and \emph{temporal click models} that consider actual click orders observed in user interactions, and thus allow for non-sequential behaviors such as revisiting.
Additionally, \citeauthor{Liu2024} introduce a separate taxonomy for \gls{nn} click models, thereby implicitly indicating that conceptually \gls{pgm} and \gls{nn} click models should not be compared directly.

\subsection{Click models beyond single-list layouts}
Compared to the single-list layout that is common in web search, research has paid considerably less attention to car\-ou\-sels or grid layouts. 
Nevertheless, carousel interfaces have become the \emph{de facto} standard for modern streaming services \cite{10.3389/fdata.2023.1239705,10.1145/3383313.3412217, 10.1145/2959100.2959174, 10.1145/3289600.3291027}. %
Carousel layouts have multiple lists, each with a topic or theme displayed as a header that defines and groups the items in that carousel/list \cite{10.1145/3452918.3465493, 10.1145/3640544.3645223, 10.1145/3604915.3610638}.
Although grid layouts also distribute items horizontally and vertically,
they do not group items into multiple lists, but just a single collection, and thus do not present clusters of items or item themes~\cite{10.1145/1743666.1743736, doi:10.1080/10447318.2013.846790, 10.1145/2959100.2959150}. Moreover, carousel interfaces allow for horizontal scrolling of each list  (\textit{swiping}) while grids do not.

There is very little research on carousel interfaces~\cite{10.3389/fdata.2023.1239705, 10.1145/3383313.3412217, 10.1145/3450614.3461680, 10.1145/3604915.3610638}, especially related to click models.
To the best of our knowledge, there are only two published papers on interaction/click models for carousel interfaces:
\citet{10.1145/3511095.3531278} assume that users keep browsing topics until they find the desired one, and then browse only the items within that topic until they find the desired item.
Based on this assumption, they propose the \gls{ccm1}~\cite{10.1145/3643709}, which can be seen as a variant of the cascade model in which users may leave unsatisfied if they do not find an attractive topic or item.
Compared with the wide variety of click models for web search interfaces~\citep{chuklin_markov_rijke_click_models_2015}, the proposed carousel click model appears simplistic~\citep{ 10.1145/3643709},
despite the fact that carousel interfaces allow for more complex user behavior.
Thus, click modeling for carousel interfaces remains a barely explored open research area.

Click models for grid layouts have received more attention~\citep{10.1145/3077136.3080799, 10.1145/3209978.3209990, 10.1145/3442381.3450098, 10.1145/3394486.3403336, 10.1145/2959100.2959150, 10.1145/3437963.3441749, 10.1145/1935826.1935873}.
\citet{10.1145/3077136.3080799} study the user examination behavior in grid-based image search and found a middle position bias instead of the traditional ``F-shaped'' pattern.
Based on this finding, the \gls{gubm} was proposed as an interactive behavior model for web image search~\citep{10.1145/3209978.3209990}.
\citet{10.1145/3442381.3450098} use neural networks to build a click model that avoids restrictive assumptions about the user interface or user behavior; and thereby, it is able to capture a variety of user behavior patterns in grid layouts.

\section{Preliminaries}
\label{sec:prelim}

In this section we introduce our notation and define key concepts that are used in the remainder of this paper.

Carousel interfaces display items in multiple horizontal lists; each has a title that represents their theme or topic.
We use the tuple $T$ to represent the topics of the $M$ displayed lists; for brevity, we also refer to \emph{themes} as \emph{topics}:
\begin{equation}
    T = (T_1, T_2, \dots, T_M).
\end{equation}
Thus, $T_1$ refers to the topic of the first displayed list.

For ease of notation, we assume that every list consists of $N$ items.
To denote the displayed items and their locations, we use the tuple $Y$ with two indices per item:
\begin{equation}
    Y = (Y_{1,1}, Y_{1,2}, \ldots Y_{1,N}, Y_{2,1}, Y_{2,2}, \dots, Y_{M,N}).
\end{equation}
Equivalently, $Y$ can be treated as an $M \times N$ matrix containing item-ids.
In either case, $Y_{i,j}$ indicates the $j$-th item in the $i$-th list (corresponding to topic $T_i$).
Furthermore, clicks are denoted in a similar manner to match the position of their corresponding items:
\begin{equation}
    C = (C_{1,1}, C_{1,2}, \ldots C_{1,N}, C_{2,1}, C_{2,2}, \dots, C_{M,N}).
\end{equation}
Thus, $C_{i,j} = 1$ indicates a click on item $Y_{i,j}$ displayed under topic $T_i$.
For many of our equations, we need to denote a dependency between all variables in $C$ except a single click variable $C_{i,j}$.
To keep our notation brief, we introduce the following variable for this purpose:
\begin{equation}
    C' = C \setminus C_{i,j},
\end{equation}
where $i$ and $j$ should be clear from the surrounding context.

Finally, our discussion also needs notation to discuss single-list and grid interfaces.
For simplicity and brevity, our notation treats single-list interfaces as carousel interfaces with a single list and the first indices and topic are ignored.
In other words, if $Y$ represents a single list of $N$ items then $Y_i$ is the $i$-th item in the list, the same goes for clicks $C$.
Similarly, we treat grid interfaces as carousel interfaces by ignoring topic variables.

We note that scrolling actions of users (vertical or horizontal) can be logged in some applications, and thus, could also be used by click models.
However, as these actions are not ubiquitously recorded and traditionally not included in click models, we leave their inclusion for future work.

\section{Click Model Taxonomies Should Firstly Concern Models, Not Behaviors}
\label{sec:per1}
The main goal of this work is to reconsider what mathematical properties should be the fundamental concepts in click model design.
More specifically, we aim to provide an overview of all possible click model designs, by providing a taxonomy for the categorization of click models.
In order for this taxonomy to be relevant and applicable to the click models of past and future work, we have three principle desiderata:
\begin{itemize}[leftmargin=*]
\item \emph{\textbf{Comprehensiveness:}} \emph{The taxonomy should be \textbf{comprehensive} such that it is applicable to all possible click models.}
\item \emph{\textbf{Exclusivity:}} \emph{The taxonomy should be \textbf{exclusive} such that each click model falls into a single category without overlaps.}
\item \emph{\textbf{Stability:}} \emph{The taxonomy should be \textbf{stable} such that it avoids future structural changes as new click models emerge.}
\end{itemize}
As a result, we arrive at the following position:
\begin{enumerate}[leftmargin=0em]
\item[] \emph{Click model taxonomies should primarily categorize based on the modeled relationships between observed variables; assumed user behavior or latent variables should not be a main concern.}
\end{enumerate}
We believe our position constitutes a new perspective in click model categorization, as existing work has taken the opposite approach: i.e., \gls{pgm} click models have been categorized based on their underlying assumed user behaviors~\citep{chuklin_markov_rijke_click_models_2015, Liu2024};
and generally, \gls{nn} click models have been characterized by their network structure, which are their latent variables \cite{10.1145/2872427.2883033, 10.1145/3626772.3657939, 10.1145/3442381.3450098, 10.1145/3336191.3371819, Liu2024}.

To be clear, we are \emph{not} arguing that user behavior assumptions or latent variables are of \emph{no} importance. Instead, we argue that a future-proof taxonomy of click models should first-and-foremost be based on how they model the relationships between observed variables.
Our position equally applies to all \textbf{single-list}, \textbf{grid} and \textbf{carousel} click models.
The remainder of this section builds the case for our position by showing that it follows as a reasonable conclusion from our three desiderata.

\subsection{The case against user behavior assumptions as a taxonomy basis} 
\label{sec:behaviorassumptions}
The earliest click models were designed as \glspl{pgm} (or can be reformulated as such). As a result, their designs were empirical and thoughtfully hand-crafted~\citep{chuklin_markov_rijke_click_models_2015}.
The introduction of new models was often inspired by the inability of previous models to capture certain user behaviors.
Consequently, most \gls{pgm} click models are designed for specific user behavior and are categorized accordingly in previous work~\citep{chuklin_markov_rijke_click_models_2015, Liu2024}.
This trend has changed with the introduction of \gls{nn}-based click models, since \glspl{nn} require little handcrafting, as their expressiveness enables them to capture almost any arbitrary user behavior.

We argue that using user behavior assumptions as the basis of our taxonomy goes against all three of our desiderata.

The issue with \emph{comprehensiveness} is that there appears to be an endless number of possible assumptions one can make about user behavior, since one can conceptualize countless behaviors.
For instance, examination can modeled by one binary variable, but one could also distinguish between \emph{careful (read)} and \emph{skim} examinations \cite{10.1145/2661829.2661907} through two binary variables, or two probability variables, etc.
It appears impossible for a taxonomy to account for all possible assumptions, yet, to be comprehensive, it should know how to categorize them all.
Therefore, user behavior assumptions seem infeasible as a basis for a comprehensive taxonomy.

The problem with \emph{exclusivity} and \emph{stability} is more complex.
Our argument is based on the fact that different user behaviors can result in mathematically-equivalent click models.
To illustrate this, we consider two example of user behaviors.
First is the well-known \gls{pbm} \cite{10.1145/1242572.1242643} that factorizes a click probability as a product of an examination and relevance probabilities:
\begin{equation}
\begin{split}
&P(\text{click} \mid \text{position}, \text{item})
\\[-0.6ex]&\quad 
= P(\text{examination} \mid \text{position})\times P(\text{relevance} \mid\text{item}).
\label{pbm}
\end{split}
\end{equation}
Second is a click model based on trust bias~\cite{10.1145/1076034.1076063} where all items are examined equally.
Decisions to click are based on how relevant items are perceived to be, with an additional trust factor determined by where an item is ranked, i.e., the user trusts the system and thus highly ranked items are more likely to be clicked:
\begin{equation}
\begin{split}
&P(\text{click} \mid \text{position}, \text{item})
\\[-0.6ex]&\;\;
= P(\text{trust} \mid \text{position})\times P(\text{perceived relevance} \mid\text{item}).
\end{split}
\label{eq:second}
\end{equation}
In terms of user behavior, there is an enormous difference: \eqref{pbm} captures variability in examination, \eqref{eq:second} does not, and \eqref{eq:second} instead differentiates trust and perceived relevance.
Thus, if user behavior assumptions were the basis for categorization, it is unlikely that these models would belong to the same category.
However, in mathematical terms, these models are equivalent: both describe a product of a position-factor with an item-factor.
Thus, for modeling click probabilities, these models can be used interchangeably.
With a taxonomy based on user behavior assumptions, this contrast poses a problem for \emph{exclusivity} as mathematically equivalent models could be categorized differently.
Additionally, it also poses a \emph{stability} problem: when equivalent models can be proposed that assume substantially different behaviors, new categories become necessary.
As a result, these taxonomies may require continual expansion rather than remain stable.

Finally, we note that it is \emph{not} our goal to create a categorization of \emph{user behavior}, but to create a taxonomy of \emph{click models}.
Whilst there is value in both, these are different aims with different purposes and should be treated accordingly.
Hence, we conclude that user behavior assumptions should not form the fundamental basis of the categorization of click models.

\subsection{The case against using latent variables in click model taxonomies} 
\label{sec:againstlatent}

Apart from trivial baselines, all \gls{pgm}-based click models use latent variables.
Based on this common usage, previous categorizations of \gls{pgm} click models have assumed the presence of certain latent variables; in particular variables representing \emph{examination}, \emph{attractiveness}, and \emph{user satisfaction}~\citep{chuklin_markov_rijke_click_models_2015, Liu2024}.

We argue that one should avoid using latent variables in click model taxonomies, in order to meet our three desiderata.

First of all, we point out that latent variables come with implicit assumptions about user behavior.
For instance, treating \emph{examination}, \emph{attractiveness} or \emph{satisfaction} as binary variables assumes that they are discrete events that cannot be nuanced beyond one-dimensional probabilities.
Although the actual concepts behind these variables are complex, multifaceted, and equivocal \cite{10.1145/2661829.2661907, chuklin_markov_rijke_click_models_2015}, in order to be useful, models have to simplify the concepts they represent, we argue that a future-proof taxonomy should not prescribe such conceptualizations.
Moreover, this also raises the problems that assumptions about user behavior pose for our desiderata as discussed in Section~\ref{sec:behaviorassumptions}.

Furthermore, it is unclear how \gls{nn}-based click models should be treated in categorizations based on latent variables.
Whilst hybrid click models exist that combine \gls{pgm} and \glspl{nn}~\citep{10.1145/3159652.3159732}\footnote{\citet{10.1145/3159652.3159732} combine \gls{pgm} and \gls{gbdt}; a \gls{pgm} and \glspl{nn} version of this click model is provided by \citet{10.1145/3292500.3330677}.}
and other \gls{nn} click models have a connection to \gls{pgm} counterparts~\citep{10.1145/3442381.3450098},
many \gls{nn} click models are not intended to model latent variables that represent user behavior~\citep{10.1145/2872427.2883033, 10.1145/3626772.3657939, 10.1145/3336191.3371819}.
Technically speaking, the hidden layers of the \glspl{nn} could be considered latent variables.
However, we argue that categorizations using latent variables cannot capture many important similarities between \glspl{nn} and \glspl{pgm}, and instead, that categorizations should use aspects that naturally apply to both model types.

\subsection{The case for the relationships between observed variables as a taxonomy basis}

To avoid the aforementioned limitations, we propose that click model taxonomies should firstly be based the mathematical relations between clicks and the other observable variables.
In our setting, the observed variables are the clicks $C$, the displayed items $Y$ and topics $T$, in other settings, there may be additional observed variables.
The mathematical relations between the observed variables are captured by how the conditional distribution $P(C \mid Y, T)$ is modeled, for instance, whether some variables are independent (e.g., $i\not=j \rightarrow P(C_i) \perp\!\!\!\perp P(C_j)$) or correlated in a certain way (e.g., $(C_i = 1 \land j > i) \rightarrow P(C_j) = 0$).
We argue that a taxonomy based on the mathematical relations between observed variables satisfies \emph{all} three of our desiderata, whereas previous categorizations \emph{do not} (as laid out in Section~\ref{sec:behaviorassumptions} and~\ref{sec:againstlatent}).

Considering \emph{comprehensiveness}, a taxonomy based on the relations between observed variables can encompass \emph{all} possible models.
Firstly, consider that every click model must define a relationship with all observed variables, either implicitly or explicitly; models that ignore certain observed variables are actually implicitly modeling an independent relationship between these variables and clicks (e.g., in the \gls{rcm}~\cite{chuklin_markov_rijke_click_models_2015} $C_i \perp\!\!\!\perp (C', Y, T)$).
Therefore, the categorization criteria of such taxonomies can be applied to any existing or future click model.
Secondly, these criteria are agnostic to the inner workings of models and thus apply to any type of model, i.e., \gls{pgm}, \gls{nn}, \gls{gbdt}, or any potential new types.
Accordingly, we argue that taxonomies should first categorize on what relations between observed variables a click model can capture, not on the technical details of how the models function (cf.\ the Markov blanket of \citet{pearl2014probabilistic}).

Another key advantage of this approach is it can easily guarantee \emph{exclusivity} between mathematically equivalent click models. 
This is the case because criteria regarding mathematical relationships between observed variables can guarantee non-overlapping categories with ease, i.e., categories based on whether a model meets a logical criteria or not.
This is in heavy contrast with categories based on latent variables or assumptions about user behaviors, and avoids the situations laid out in Section~\ref{sec:behaviorassumptions}.
Therefore, since it is straightforward to choose non-overlapping categorization criteria in this approach, it can easily guarantee \emph{exclusivity} in click model taxonomies.

Finally, we argue that our approach also results in great \emph{stability} for resulting taxonomies.
We build on the arguments laid out in the previous paragraphs.
Firstly, because our approach is agnostic to the inner structure of the models, it does not have to change when a new type of model is introduced.
For instance, if a future novel click model is based on a hypothetical successor to the \gls{nn}, it can still be categorized by the taxonomies of our approach.
Secondly, similarly, the introduction of new observed variables would only require an expansion of a taxonomy, not a change in structure of the existing part.
As stated before, models that ignore a certain variable are actually implicitly assuming that clicks are conditionally independent of that variable; thus, for any new variable $Z$ introduced, all existing models assume $P(C \mid T, Y, C', Z) = P(C \mid T, Y, C')$.
One could make a split in the taxonomy between models that assume independence with the new variable and those that do not, the existing taxonomy would then fall under the first part of the split.
In fact, this is what our taxonomy does with the topic variable $T$; if one removes the \emph{Topic-Dependent} sub-graph, the remainder of the taxonomy still functions as a standalone taxonomy of click models for single-list and grid interfaces (see Section~\ref{sec:tax_carousel} and Figure~\ref{fig:tree_structure}).
Therefore, because the introduction of new model types or new observed variables would not require an overhaul of an existing taxonomy, we argue that our approach has \emph{stability} by providing a taxonomy that one can expect to remain consistent over time.

In conclusion, we propose that click model taxonomies should be based firstly on how they model the relationships between clicks and the other observed variables. 
This approach returns to the essence of click modeling, that is, modeling previous clicks to predict future click probabilities, rather than trying to propose a new model for every potential user behavior.
Finally, to be clear, we are \emph{not} arguing that other properties of click models do not matter, e.g., latent variables can be important for explainability \cite{chuklin_markov_rijke_click_models_2015, 10.1145/1341531.1341545} or propensity estimation \cite{10.1145/3018661.3018699,10.1145/3437963.3441794}, and \gls{nn}-based click models often provide better predictive performance \cite{10.1145/2872427.2883033, 10.1145/3209978.3210004}.
However, our position is that the mathematical relationships between observed variables should be the first consideration when categorizing, designing or comparing click models, the consideration of other properties should be secondary.

\section{Three Key Design Choices of Click Models} \label{sec:choice}

Traditional click models designs are generally motivated by specific user behavior to be captured~\citep{chuklin_markov_rijke_click_models_2015}, which thus serves as a starting point for the design process.

In contrast, and in line with our position, we argue that the relationships between observed variables should be the first consideration for click model design.
More specifically, we pose that \textbf{three} key design choices must be answered, and that these will define how observed variables interact with and influence click events.
In order, these three decisions are:
\begin{enumerate}[leftmargin=*]
\item \textbf{Global dependencies:} \emph{On what collections of observed variables are click probabilities conditioned?}
\item \textbf{Sequentiality:} \emph{On what subsequences of these collections are individual click probabilities conditioned?}
\item \textbf{Factorization:} \emph{How is the influence of the conditional variables on individual click probabilities modeled?}
\end{enumerate}
Together, these decisions determine how a click model estimates click probabilities conditioned on the observable information.
Generally, they make a trade-off between the expressiveness of the model, i.e., how many patterns it can model, and its complexity, i.e., the number of parameters and proneness to overfitting of the model.
The remainder of this section details each decision and describes potential options for each.

\subsection{Global dependencies}
The first decision to make is to select which collections of observable variables could have a direct influence on click events.
In the case of carousel interfaces, the possible variable tuples are the topics $T$, the items $Y$ and the other clicks $C'$ (excluding self-dependencies).
In one extreme case, clicks are independent from all variables: $P(C \mid T, Y, C') = P(C)$, or for example, one can choose to model clicks as independent of each other: $P(C \mid T, Y, C') = P(C \mid T, Y)$.
Therefore, the options for this choice are any subset of the tuples of observable variables; formally, in our setting, one must choose a set $X \subseteq \{T, Y, C'\}$ which results in the click model assumption:
\begin{equation}
P(C \mid T, Y, C') = P(C \mid X).
\end{equation}
We note that every click model must assume an answer to this decision, implicitly or explicitly.

\subsection{Sequentiality}
Global dependencies describe relations between the collections or tuples of variables $T$, $Y$ and $C$.
However, these dependencies may be too broad, that is, one may not want to model every item in $Y$ to influence every click in $C$.
Instead, many existing click models only model a dependency between parts of the tuples, which decreases their complexity.

Accordingly, the second decision is to decide on which parts of the tuples individual click probabilities should be conditioned.
Naturally, variables excluded by the previous decision about global dependencies can not be chosen for this subsequent decision.
For example, one can decide that click probabilities should only be conditioned by aspects of the clicked item themselves:
\begin{equation}
P(C_{i,j} \mid T, Y, C') = P(C_{i,j} \mid T_{i}, Y_{i,j}),
\end{equation}
or alternatively, in a single-list setting, on all preceding items and clicks, plus the item itself:
\begin{equation}
P(C_i \mid Y, C') = P(C_i \mid Y_{1:i}, C'_{1:i-1}).
\end{equation}

In order to produce a mathematically valid model from which predictions can be computed, one should avoid cyclical dependencies. 
For example, if in a single-list setting one states that clicks depend on the preceding click:
\begin{equation}
i > 1 \rightarrow P(C_i \mid Y, C') = P(C_i \mid C'_{i-1}),
\end{equation}
but also states that the first click depends on the last click:
\begin{equation}
P(C_1 \mid C'_{N}),
\end{equation}
then a cyclical dependency is created where a predicted click probability indirectly depends on the click itself.
One should avoid such decisions as they can make the computability of model predictions intractable.
Hence, we argue that one should choose a sequence of (groups of) items that describe a chain of conditionals in the form of a directed acyclic graph \cite{10.5555/1162264}.

Importantly, this decision is about the dependencies between observed variables, it is not (directly) about assuming a sequence by which a user examines items.
Nevertheless, there is a possible correspondence or correlation between the two, for instance, if click probabilities are conditioned on preceding items that may be because one assumes that a user may examine these items first.
But it does not have to be the case that the conditionals describe which items a users considers when deciding to click, there may be other reasons that lead to an observable dependency.
Again, our first consideration is the modeled relations between variables, not what user behaviors they attempt to describe.

\subsection{Factorization}

The previous decisions have determined what variables are used as conditionals for click probabilities in our model, the final question is how to transform the variable values into a click probability.
In other words, how is the relationship between a click probability and its conditional variables modeled.
In line with our position, we argue that this decision should be made without considering latent variables or what modeling technique is used as a basis (i.e., \gls{pgm} or \gls{nn}).
Instead, we propose that the model should be described in terms of functions on the observed variables and simplified as much as possible.
This can reveal mathematical equivalence between models and precisely describe the patterns that they can capture.

For example, the models described in (\ref{pbm}) and (\ref{eq:second}) provide very different narratives to how clicks come to be, and assume different latent variables in the click process.
Yet, if we were to describe these models as mathematical functions according to our factorization decision approach, they reveal the same product of a position factor and an item factor:
\begin{equation}
P(\text{click} \mid \text{position}, \text{item}) = f(\text{position})\times g(\text{item}).
\label{eq:factorizationexample}
\end{equation}
Thereby revealing that they model the same mathematical relations between clicks and observed variables, and thus, that they have identical expressiveness.

By considering factorization, we can not only compare models that use different latent variables or describe different user behaviors, it also enables us to compare different model types.
For example, the traditional \gls{pgm}-based \gls{pbm} \cite{10.1145/1242572.1242643} and the later Regression-EM \gls{pbm} \cite{10.1145/3159652.3159732} can be factorized in identically, but whilst the former is purely a \gls{pgm}, the latter uses regression by a \gls{gbdt} to estimate the item factor.

Some \gls{nn}-based click models are explicitly designed around factorization in order to disentangle the effect of presentation biases from user preferences.
For instance, two-tower models explicitly process information about other items and their positions separately from features of the item itself up until the final step of the model~\citep{10.1145/3442381.3450098, 10.1145/3477495.3531837, 10.1145/3580305.3599914}.

Thus with the final decision of factorization, we argue that the most salient mathematical properties of a click model are described, to allow for click model categorization and comparison.
The following sections discusses how these three decisions can be used to create a taxonomy of existing and future click models and help design  new click models.

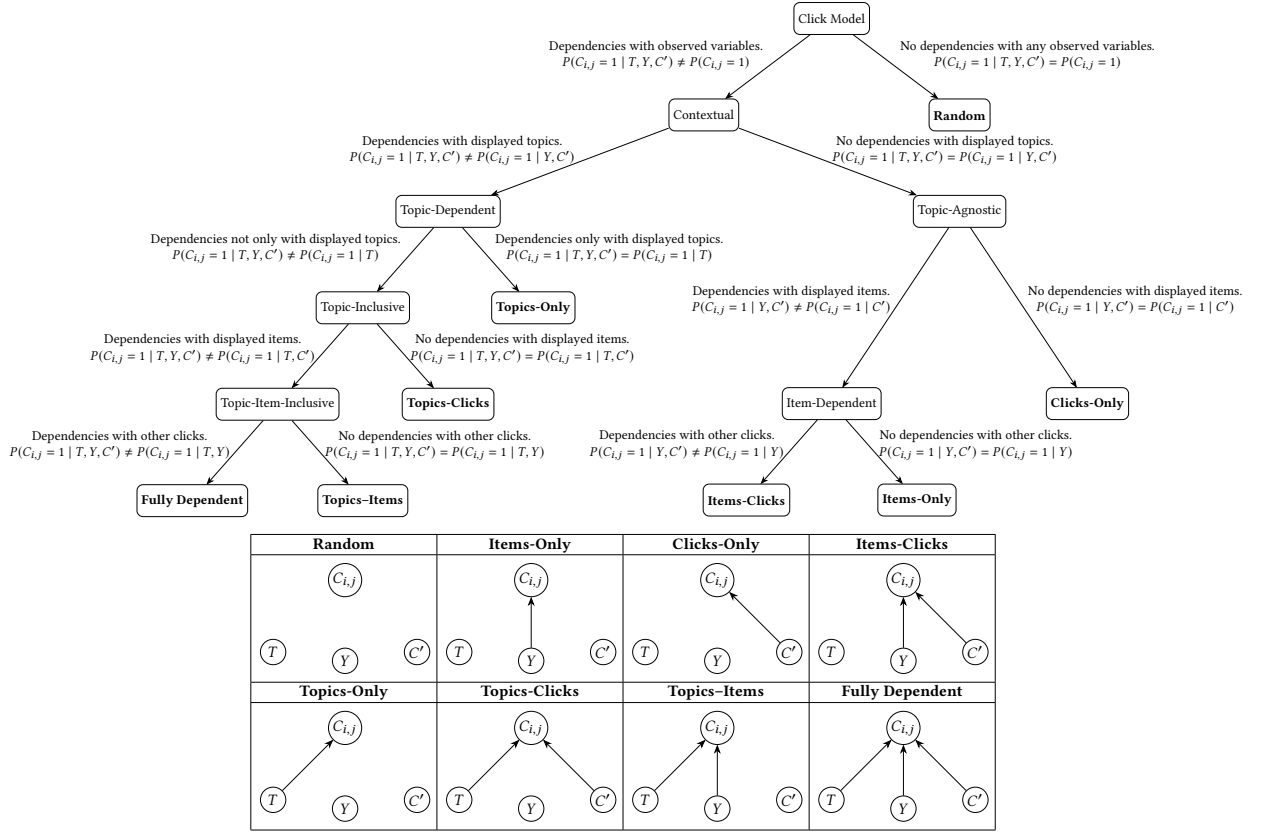
\begin{figure*}[ht]
    \centering
    \begin{adjustbox}{width=\textwidth}
    \begin{tikzpicture}[
    transform shape,
  every node/.style={draw, rectangle, rounded corners, minimum height=0.75cm},
  every path/.style={-Stealth},%
  label/.style={draw=none, rectangle} %
]

\node (A) {Click Model};
\node (B) [below=1.5cm of A, xshift=-3cm] {Contextual};
\node (C) [below=1.5cm of A, xshift=3cm] {\textbf{Random}};
\node (D) [below=1.5cm of B, xshift=-6cm] {Topic-Dependent};
\node (E) [below=1.5cm of B, xshift=6cm] {Topic-Agnostic};
\node (F) [below=3.75cm of E, xshift=3cm] {\textbf{Clicks-Only}}; %
\node (G) [below=3.75cm of E, xshift=-3cm] {Item-Dependent};       %
\node (H) [below=1.5cm of D, xshift=-2cm] {Topic-Inclusive};
\node (I) [below=1.5cm of D, xshift=2cm] {\textbf{Topics-Only}};
\node (J) [below=1.5cm of H, xshift=2cm] {\textbf{Topics-Clicks}};
\node (K) [below=1.5cm of H, xshift=-2cm] {Topic-Item-Inclusive};
\node (L) [below=1.5cm of K, xshift=-2cm] {\textbf{Fully Dependent}};
\node (M) [below=1.5cm of K, xshift=2cm] {\textbf{Topics–Items}};
\node (N) [below=1.5cm of G, xshift=-2cm] {\textbf{Items-Clicks}};
\node (O) [below=1.5cm of G, xshift=2cm] {\textbf{Items-Only}};

\draw (A) -- node[label, above left, align=center, yshift=-0.2cm] {Dependencies with observed variables. \\ $P(C_{i,j}=1 \mid T, Y, C') \neq P(C_{i,j}=1)$} (B);
\draw (A) -- node[label, above right, align=center, yshift=-0.2cm] {No dependencies with any observed variables. \\ $P(C_{i,j}=1 \mid T, Y, C') = P(C_{i,j}=1)$} (C);
\draw (B) -- node[label, above left, align=center, yshift=-0.2cm] {Dependencies with displayed topics. \\ $P(C_{i,j}=1 \mid T, Y, C') \neq P(C_{i,j}=1 \mid Y, C')$} (D);
\draw (B) -- node[label, above right, align=center, yshift=-0.2cm] {No dependencies with displayed topics. \\$P(C_{i,j}=1 \mid T, Y, C') = P(C_{i,j}=1 \mid Y, C')$} (E);
\draw (E) -- node[label, above right, align=center, yshift=-0.3cm] {No dependencies with displayed items. \\ \(P(C_{i,j}=1 \mid Y, C') = P(C_{i,j}=1 \mid C')\)} (F);
\draw (E) -- node[label, above left, align=center, yshift=-0.3cm] {Dependencies with displayed items. \\ \(P(C_{i,j}=1 \mid Y, C') \neq P(C_{i,j}=1 \mid C')\)} (G);
\draw (D) -- node[label, above left, align=center, yshift=-0.2cm] {Dependencies not only with displayed topics. \\ $P(C_{i,j}=1 \mid T, Y, C') \neq P(C_{i,j}=1 \mid T)$} (H);
\draw (D) -- node[label, above right, align=center, yshift=-0.2cm] {Dependencies only with displayed topics. \\ $P(C_{i,j}=1 \mid T, Y, C') = P(C_{i,j}=1 \mid T)$} (I);
\draw (H) -- node[label, above right, align=center, yshift=-0.3cm] {No dependencies with displayed items. \\  \(P(C_{i,j}=1 \mid T,Y,C') = P(C_{i,j}=1 \mid T,C')\)} (J);
\draw (H) -- node[label, above left, align=center, yshift=-0.3cm] {Dependencies with displayed items. \\  \(P(C_{i,j}=1 \mid T,Y,C') \neq P(C_{i,j}=1 \mid T,C')\)} (K);
\draw (K) -- node[label, above left, align=center, yshift=-0.3cm] {Dependencies with other clicks. \\  \(P(C_{i,j}=1 \mid T,Y,C') \neq P(C_{i,j}=1 \mid T,Y)\)} (L);
\draw (K) -- node[label, above right, align=center, yshift=-0.3cm] {No dependencies with other clicks. \\ \(P(C_{i,j}=1 \mid T,Y,C') = P(C_{i,j}=1 \mid T,Y)\)} (M);
\draw (G) -- node[label, above left, align=center, yshift=-0.3cm] {Dependencies with other clicks. \\ \(P(C_{i,j}=1 \mid Y,C') \neq P(C_{i,j}=1 \mid Y)\)} (N);
\draw (G) -- node[label, above right, align=center, yshift=-0.3cm] {No dependencies with other clicks. \\ \(P(C_{i,j}=1 \mid Y,C') = P(C_{i,j}=1 \mid Y)\)} (O);

\end{tikzpicture}
\end{adjustbox}
\\
\vspace{0.5\baselineskip}
\begin{adjustbox}{width=0.6\textwidth}
\begin{tabular}{| c | c | c | c |}
\hline
\textbf{Random} & \textbf{Items-Only} & \textbf{Clicks-Only} & \textbf{Items-Clicks} \\ \hline
\begin{minipage}[c][2.5cm][c]{0.2\textwidth}
\centering
\begin{tikzpicture}[baseline=-0.5ex,
  level distance=8mm, sibling distance=8mm,
  every node/.style={draw,circle, inner sep=1pt, minimum size=0.5cm},
  every path/.style={-Stealth}
  ]
\node (A) {$C_{i,j}$};
\node (B) [below left=of A] {$T$};
\node (C) [below=of A] {$Y$};
\node (D) [below right=of A] {$C'$};
\end{tikzpicture}
\par
\end{minipage}

&

\begin{minipage}[c][2.5cm][c]{0.2\textwidth}
\centering
\begin{tikzpicture}[baseline=-0.5ex,
  level distance=8mm, sibling distance=8mm,
  every node/.style={draw,circle, inner sep=1pt, minimum size=0.5cm},
  every path/.style={-Stealth}]
\node (A) {$C_{i,j}$};
\node (B) [below left=of A] {$T$};
\node (C) [below=of A] {$Y$};
\node (D) [below right=of A] {$C'$};

\draw (C) -- (A);

\end{tikzpicture}
\end{minipage}
&

\begin{minipage}[c][2.5cm][c]{0.2\textwidth}
\centering
\begin{tikzpicture}[baseline=-0.5ex,
  level distance=8mm, sibling distance=8mm,
  every node/.style={draw,circle, inner sep=1pt, minimum size=0.5cm},
  every path/.style={-Stealth}]
\node (A) {$C_{i,j}$};
\node (B) [below left=of A] {$T$};
\node (C) [below=of A] {$Y$};
\node (D) [below right=of A] {$C'$};

\draw (D) -- (A);

\end{tikzpicture}
\end{minipage}
&

\begin{minipage}[c][2.5cm][c]{0.2\textwidth}
\centering
\begin{tikzpicture}[baseline=-0.5ex,
  level distance=8mm, sibling distance=8mm,
  every node/.style={draw,circle, inner sep=1pt, minimum size=0.5cm},
  every path/.style={-Stealth}]
\node (A) {$C_{i,j}$};
\node (B) [below left=of A] {$T$};
\node (C) [below=of A] {$Y$};
\node (D) [below right=of A] {$C'$};

\draw (C) -- (A);
\draw (D) -- (A);

\end{tikzpicture}
\end{minipage}
\\ \hline
\textbf{Topics-Only} & \textbf{Topics-Clicks} & \textbf{Topics–Items} & \textbf{Fully Dependent} \\ \hline
\begin{minipage}[c][2.5cm][c]{0.2\textwidth}
\centering
\begin{tikzpicture}[baseline=-0.5ex,
  level distance=8mm, sibling distance=8mm,
  every node/.style={draw,circle, inner sep=1pt, minimum size=0.5cm},
  every path/.style={-Stealth}]
\node (A) {$C_{i,j}$};
\node (B) [below left=of A] {$T$};
\node (C) [below=of A] {$Y$};
\node (D) [below right=of A] {$C'$};

\draw (B) -- (A);

\end{tikzpicture}
\end{minipage}
&
\begin{minipage}[c][2.5cm][c]{0.2\textwidth}
\centering
\begin{tikzpicture}[baseline=-0.5ex,
  level distance=8mm, sibling distance=8mm,
  every node/.style={draw,circle, inner sep=1pt, minimum size=0.5cm},
  every path/.style={-Stealth}]
\node (A) {$C_{i,j}$};
\node (B) [below left=of A] {$T$};
\node (C) [below=of A] {$Y$};
\node (D) [below right=of A] {$C'$};

\draw (B) -- (A);
\draw (D) -- (A);

\end{tikzpicture}
\end{minipage}
&

\begin{minipage}[c][2.5cm][c]{0.2\textwidth}
\centering
\begin{tikzpicture}[baseline=-0.5ex,
  level distance=8mm, sibling distance=8mm,
  every node/.style={draw,circle, inner sep=1pt, minimum size=0.5cm},
  every path/.style={-Stealth}]
\node (A) {$C_{i,j}$};
\node (B) [below left=of A] {$T$};
\node (C) [below=of A] {$Y$};
\node (D) [below right=of A] {$C'$};

\draw (B) -- (A);
\draw (C) -- (A);
\end{tikzpicture}
\end{minipage}
&

\begin{minipage}[c][2.5cm][c]{0.2\textwidth}
\centering
\begin{tikzpicture}[baseline=-0.5ex,
  level distance=8mm, sibling distance=8mm,
  every node/.style={draw,circle, inner sep=1pt, minimum size=0.5cm},
  every path/.style={-Stealth}]
\node (A) {$C_{i,j}$};
\node (B) [below left=of A] {$T$};
\node (C) [below=of A] {$Y$};
\node (D) [below right=of A] {$C'$};

\draw (B) -- (A);
\draw (C) -- (A);
\draw (D) -- (A);

\end{tikzpicture}
\end{minipage}
\\ \hline
\end{tabular}
\end{adjustbox}
\vspace*{-0.75\baselineskip} 
    \caption{Overview of part of our proposed click model taxonomy based on the \emph{global dependencies} decision.
    We further categorize based on \emph{sequentiality} and \emph{factorization}; not displayed here to keep the overview brief and simple.}
    \label{fig:tree_structure}
\vspace*{-0.75\baselineskip}    
\end{figure*}

\section{Theory-Based Taxonomy of Click Models}
\label{sec:tax_carousel}
In this section, building on our three key design choices, we create a taxonomy for single-list, grid, and carousel click models, grounded in how they model the relationships between clicks and the other observed variables.
Our taxonomy first categorizes click models based on their decisions on global dependencies, i.e., on which subset of variable tuples the click probability conditioned.
There are eight possible subsets of observed variable tuples which can be chosen as the conditionals of click probabilities:
\begin{equation}
\{ \emptyset, \{T\}, 
\{Y\}, \{C'\}, \{T, Y\}, \{T, C'\}, \{Y, C'\}, 
\{T, Y, C'\} \}.
\end{equation}
We note that for the single-list and grid interfaces, it does not make sense to choice sets with the topic $T$ tuple since these are not present in those interfaces, leaving four choices of subsets.

Thus, the first design choice results in eight categories, Figure~\ref{fig:tree_structure} provides an overview, we briefly describe each here:
\begin{itemize}[leftmargin=*]
    \item \textbf{Random.}
    The click probability is constant, independent of the displayed items, clicks on other items, or the topic.
    Applicable to all interfaces.

    \item \textbf{Clicks-Only.}     
    Click probabilities are conditioned on other clicks, but not on what items or topics are displayed.
    Applicable to all interfaces.

    \item \textbf{Items-Only.}
    These models capture the effect of the displayed items on clicks, e.g., they can model item relevance or popularity, but ignore topics and assume no dependency between clicks.
    Applicable to all interfaces.

    \item \textbf{Items-Clicks.}
    Click probabilities are conditioned on displayed items and other clicks, but not on topics.
    These models can capture item relevance or popularity and cannibalization effects between clicks.
    Applicable to all interfaces.

    \item \textbf{Topics-Only.}
    Modeled click behavior is determined solely by the displayed topics, ignoring influences from  items or other clicks.
    This assumes only a carousel's topic matters for interactions with it.
    Only relevant for carousel interfaces.

    \item \textbf{Topics-Clicks.}
    These models condition both on the displayed topics and other clicks, but not on displayed items.
    Thus, they assume clicks occur in interdependent patterns that depend on topics.
    Only relevant for carousel interfaces.
    
    \item \textbf{Topics-Items.}
    Click probabilities are conditioned on both displayed topics and items, but not on other clicks.
    These models capture dependencies with topics and items but assume independence between clicks (once conditioned on topics and items).
    Relevant only for carousel interfaces.

    \item \textbf{Fully Dependent.}
    These models condition on all three variable tuples: displayed topics, displayed items, and clicks on other items.
    Thereby, this category contains the most expressive click models but also the ones with the most complexity.
    Relevant only for carousel interfaces.
\end{itemize}

\noindent%
This concludes our description of the categories that follow from the first design choice concerning \textbf{global dependencies}.
We further categorize click models based on the other two decisions regarding \textbf{sequentiality}, and \textbf{factorization}.
However, as the total number of possible choices for these decisions is unclear or impractical to enumerate, we do not discuss every possible combination of choices for all three decisions.
Instead, we provide a categorization of the most prominent existing click models in Table~\ref{tab:click-model-comparison}, where we display the choices each model makes for the design decisions (explicitly or implicitly) and group models based on these choices.

To the best of our knowledge, this is the first taxonomy of single-list, grid and carousel click models that includes \glspl{pgm}, \glspl{nn} and \glspl{gbdt}.
Thereby, we provide the first meaningful comparison of existing click models across types and interfaces, and a framework for comparison with future click models through a structured theory-based categorization concerning their salient mathematical properties. 
Additionally, we hope that our approach lays the groundwork for the future development of more click models for carousel interfaces.

\begin{table*}[t]
  \centering
  \vspace{-0.5\baselineskip}
  \caption{
  The categorization of existing click models following our proposed taxonomy that categorizes based on three different design choices that describe salient mathematical properties of each model.
  Uniquely, this categorization covers \gls{pgm}, \gls{nn} and \gls{gbdt} click models for single-list, grid and carousel interfaces. %
  }
  \vspace{-\baselineskip}
  \glsresetall
  \label{tab:click-model-comparison}
  \begin{adjustbox}{width=\textwidth}
  \begin{tabular}{lccccc}
    \toprule
    \textbf{Click Models in Category} & \textbf{Global Dependencies} & & \textbf{Sequentiality} && \textbf{Factorization} \\
    \midrule
      \multirow{2}{*}{
    \begin{tabular}{l}
    \Gls*{rcm} \cite{chuklin_markov_rijke_click_models_2015}
    \end{tabular}
    } &
  \multirow{2}{*}{\begin{tabular}{c}
      Random \\
      $P(C_i = 1)$
    \end{tabular}} &
  \multirow{2}{*}{\begin{tabular}{c} \\ $=$\end{tabular}} &
  \multirow{2}{*}{\begin{tabular}{c}
      Not Applicable \\
      $P(C_i = 1)$
    \end{tabular}} &
  \multirow{2}{*}{\begin{tabular}{c} \\ $=$\end{tabular}} &
  \multirow{2}{*}{\begin{tabular}{c}
      Constant \\
      $\zeta$
    \end{tabular}} \\ \\
    \midrule
    \multirow{2}{*}{
    \begin{tabular}{l}
    \Gls*{rctr} \cite{chuklin_markov_rijke_click_models_2015}
    \end{tabular}
    } &
  \multirow{2}{*}{\begin{tabular}{c}
      Random \\
      $P(C_i=1)$
    \end{tabular}} &
  \multirow{2}{*}{\begin{tabular}{c} \\ $=$\end{tabular}} &
  \multirow{2}{*}{\begin{tabular}{c}
      Not Applicable \\
      $P(C_i=1)$
    \end{tabular}} &
  \multirow{2}{*}{\begin{tabular}{c} \\ $=$\end{tabular}} &
  \multirow{2}{*}{\begin{tabular}{c}
      Constant per Position \\
      $f(i)$
    \end{tabular}}
    \\
    \\
    \midrule
    \multirow{2}{*}{
    \begin{tabular}{l}
    \Gls*{dctr} \cite{chuklin_markov_rijke_click_models_2015}
    \end{tabular}
    } &
  \multirow{2}{*}{\begin{tabular}{c}
      Items-Only \\
      $P(C_i=1 \mid Y)$
    \end{tabular}} &
  \multirow{2}{*}{\begin{tabular}{c} \\ $=$\end{tabular}} &
  \multirow{2}{*}{\begin{tabular}{c}
      Chosen Item Only \\
      $P(C_i=1 \mid Y_{i})$
    \end{tabular}} &
  \multirow{2}{*}{\begin{tabular}{c} \\ $=$\end{tabular}} &
  \multirow{2}{*}{\begin{tabular}{c}
      Item Factor \\
      $f(Y_{i})$
    \end{tabular}} 
    \\
    \\
    \midrule
    \multirow{3}{*}{
    \begin{tabular}{l}
    \Gls*{pbm}  \\
    (Standard EM \cite{10.1145/1242572.1242643}, Regression-based EM \cite{10.1145/3292500.3330677, 10.1145/3159652.3159732})
    \end{tabular}
    } & \multirow{3}{*}{\begin{tabular}{c}
    \\ Items-Only \\ $P(C_i=1 \mid Y)$
    \end{tabular}} & 
    \multirow{3}{*}{\begin{tabular}{c} \\ \\ $=$\end{tabular}} &
    \multirow{3}{*}{\begin{tabular}{c} \\ Chosen Item Only \\ $P(C_i=1 \mid Y_{i})$
    \end{tabular}
    } & 
    \multirow{3}{*}{\begin{tabular}{c} \\ \\ $=$\end{tabular}} & \multirow{3}{*}{\begin{tabular}{c} Position Factor \\ Item Factor \\ $f(i) \cdot g(Y_{i})$
    \end{tabular}} 
    \\ \\  \\
    \midrule
    \multirow{3}{*}{
    \begin{tabular}{l}
    TrustPBM \cite{10.1145/3340531.3412031, 10.1145/3308558.3313697}
    \end{tabular}
    } &
  \multirow{3}{*}{\begin{tabular}{c}
     \\ Items-Only \\
      $P(C_i=1 \mid Y)$
    \end{tabular}} &
  \multirow{3}{*}{\begin{tabular}{c} \\ \\ $=$\end{tabular}} &
  \multirow{3}{*}{\begin{tabular}{c}
      \\ Chosen Item Only \\
      $P(C_i=1 \mid Y_{i})$
    \end{tabular}} &
  \multirow{3}{*}{\begin{tabular}{c} \\ \\ $=$\end{tabular}} &
  \multirow{3}{*}{\begin{tabular}{c}
      Two Position Factors \\ Item Factor \\ 
      $f(i) \cdot g(Y_i) + h(i)$
    \end{tabular}}
    \\
    \\
    \\
    \midrule
    \multirow{2}{*}{
    \begin{tabular}{l}
    \Gls*{csm} \cite{10.1145/3209978.3210004}
    \end{tabular}
    } & \multirow{2}{*}{\begin{tabular}{c}
    Items-Only \\ $P(C_i=1 \mid Y)$
    \end{tabular}} & 
    \multirow{2}{*}{\begin{tabular}{c} \\ $=$\end{tabular}} &
    \multirow{2}{*}{\begin{tabular}{c}  All Items \\ $ P(C_i=1 \mid Y)$
    \end{tabular}
    } & 
    \multirow{2}{*}{\begin{tabular}{c}  \\ $=$\end{tabular}} & \multirow{2}{*}{\begin{tabular}{c} Items Factor \\ $f(Y)$
    \end{tabular}} 
    \\ \\
    \midrule
    \multirow{3}{*}{
    \begin{tabular}{l}
    \Gls*{xpa} \cite{10.1145/3442381.3450098}
    \end{tabular}
    } & \multirow{3}{*}{\begin{tabular}{c}
     \\ Items-Only \\ $P(C_i=1 \mid Y)$
    \end{tabular}} & 
    \multirow{3}{*}{\begin{tabular}{c} \\ \\ $=$\end{tabular}} &
    \multirow{3}{*}{\begin{tabular}{c} \\  All Items \\ $ P(C_i=1 \mid Y)$
    \end{tabular}
    } & 
    \multirow{3}{*}{\begin{tabular}{c} \\ \\ $=$\end{tabular}} & \multirow{3}{*}{\begin{tabular}{c} Item Factor \\ Items and Position Factor \\ $f(Y_i) \cdot g(Y \setminus Y_i, i)$
    \end{tabular}}
    \\ \\  \\
    \midrule
    \multirow{2}{*}{
    \begin{tabular}{l}
    \Gls*{rbnn} \cite{10.1145/3295750.3298920}
    \end{tabular}
    } & \multirow{2}{*}{\begin{tabular}{c}
     Items-Only \\ $P(C_i=1 \mid Y)$
    \end{tabular}} & 
    \multirow{2}{*}{\begin{tabular}{c}  \\ $=$\end{tabular}} &
    \multirow{2}{*}{\begin{tabular}{c}   Items up to Chosen Item \\ $ P(C_i=1 \mid Y_{1:i})$
    \end{tabular}
    } & 
    \multirow{2}{*}{\begin{tabular}{c}  \\ $=$\end{tabular}} & \multirow{2}{*}{\begin{tabular}{c} Items Factor \\ $f(Y_{1:i})$
    \end{tabular}}
    \\ \\ 
    \midrule
    \multirow{3}{*}{
    \begin{tabular}{l}
    \Gls*{cm} \cite{10.1145/1341531.1341545}\\
    \Gls*{ubm} \cite{10.1145/1390334.1390392}\\
    \Gls*{bbm} \cite{10.1145/1557019.1557081}
    \end{tabular}
    } &
  \multirow{3}{*}{\begin{tabular}{c}
    \\  Items-Clicks \\
      $P(C_i=1 \mid Y, C')$
    \end{tabular}} &
  \multirow{3}{*}{\begin{tabular}{c}  \\ \\ $=$\end{tabular}} &
  \multirow{3}{*}{\begin{tabular}{c}
      Chosen Item \\ Clicks above Chosen Item \\
      $ P(C_i=1 \mid Y_{i}, C'_{1:i-1})$
    \end{tabular}} &
  \multirow{3}{*}{\begin{tabular}{c} \\ \\ $=$\end{tabular}} &
  \multirow{3}{*}{\begin{tabular}{c}
      Item Factor \\ Clicks Factor \\ 
      $f(Y_{i}) \cdot g\big(C'_{1:i-1}\big)$
    \end{tabular}}  \\
    \\ \\ 
    \midrule
    \multirow{3}{*}{
    \begin{tabular}{l}
    \Gls*{dcm} \cite{10.1145/1498759.1498818}
    \end{tabular}
    } &
  \multirow{3}{*}{\begin{tabular}{c}
    \\  Items-Clicks \\
      $P(C_i=1 \mid Y, C')$
    \end{tabular}} &
  \multirow{3}{*}{\begin{tabular}{c}  \\ \\ $=$\end{tabular}} &
  \multirow{3}{*}{\begin{tabular}{c}
      Chosen Item \\ Clicks above Chosen Item \\
      $ P(C_i=1 \mid Y_{i}, C'_{1:i-1})$
    \end{tabular}} &
  \multirow{3}{*}{\begin{tabular}{c} \\ \\ $=$\end{tabular}} &
  \multirow{3}{*}{\begin{tabular}{c}
      Item Factor \\ Positions-Clicks Factor \\ 
      $f(Y_{i}) \cdot g\big(1\!:\!i-1 , C'_{1:i-1}  \big)$
    \end{tabular}}  \\
    \\ \\ 
    \midrule    
    \multirow{6}{*}{
    \begin{tabular}{l}
    \Gls*{ccm2} \cite{10.1145/1526709.1526712}\\
    \Gls*{dbn} \cite{10.1145/1526709.1526711} \\
    \Gls*{pctm} \cite{10.1007/978-3-642-17537-4_40}\\
    \Gls*{pcc} \cite{10.1145/1835449.1835510} \\
    \Gls*{gcm} \cite{10.1145/1718487.1718528} \\
    \Gls*{bss} \cite{10.1145/2488388.2488508}
    \end{tabular}
    } &
  \multirow{6}{*}{\begin{tabular}{c}
    \\  Items-Clicks \\
      $P(C_i=1 \mid Y, C')$
    \end{tabular}} &
  \multirow{6}{*}{\begin{tabular}{c}  \\ \\ $=$\end{tabular}} &
  \multirow{6}{*}{\begin{tabular}{c}
      Items up to Chosen Item \\ Clicks above Chosen Item \\
      $ P(C_i=1 \mid Y_{1:i}, C'_{1:i-1})$
    \end{tabular}} &
  \multirow{6}{*}{\begin{tabular}{c} \\ \\ $=$\end{tabular}} &
  \multirow{6}{*}{\begin{tabular}{c}
      Item Factor \\ Items-Clicks Factor \\ $f(Y_{i}) \cdot g\big(Y_{1:i-1}, C'_{1:i-1}\big)$
    \end{tabular}}  \\
    \\ \\ \\ \\ \\
    \midrule
    \multirow{4}{*}{
    \begin{tabular}{l}
    \Gls*{tcm} \cite{10.1145/1835449.1835470} \\
    \Gls*{thcm} \cite{10.1145/2124295.2124334} \\
    \Gls*{pscm} \cite{10.1145/2766462.2767712} \\
    \Gls*{tacm} \cite{10.1145/2988230}
    \end{tabular}
    } & \multirow{4}{*}{\begin{tabular}{c}
    \\ Items-Clicks \\ $P(C_i=1 \mid Y, C')$
    \end{tabular}} & 
    \multirow{4}{*}{\begin{tabular}{c} \\ \\ $=$\end{tabular}} &
    \multirow{4}{*}{\begin{tabular}{c} Chosen Item \\ Clicks with Earlier Timestamps \\ $ P(C_i=1 \mid Y_i, \{c \in C' : t(c) < t(C_i)\})$
    \end{tabular}
    } & 
    \multirow{4}{*}{\begin{tabular}{c} \\ \\ $=$\end{tabular}} & \multirow{4}{*}{\begin{tabular}{c} Item Factor\\ Clicks Factor \\ $ f(Y_i) \cdot g(\{c \in C' : t(c) < t(C_i)\})$
    \end{tabular}} \\
    \\ \\ \\
    \midrule
    \multirow{3}{*}{
    \begin{tabular}{l}
    \Gls*{graphcm} \cite{10.1145/3404835.3462895} \\
    \Gls*{fetcm} \cite{10077385}
    \end{tabular}
    } & \multirow{3}{*}{\begin{tabular}{c}
     \\ Items-Clicks \\ $P(C_i=1 \mid Y,C')$
    \end{tabular}} & 
    \multirow{3}{*}{\begin{tabular}{c} \\ \\ $=$\end{tabular}} &
    \multirow{3}{*}{\begin{tabular}{c}   All Items \\ All Clicks \\ $ P(C_i=1 \mid Y,C')$
    \end{tabular}
    } & 
    \multirow{3}{*}{\begin{tabular}{c} \\ \\ $=$\end{tabular}} & \multirow{3}{*}{\begin{tabular}{c} Clicks Factor \\ Items-Clicks Factor \\ $f(C') \cdot g(Y,C')$
    \end{tabular}}
    \\ \\  \\ 
    \midrule
    \multirow{3}{*}{
    \begin{tabular}{l}
    \Gls*{ncm} \cite{10.1145/2872427.2883033}\\
    \Gls*{aicm} \cite{10.1145/3442381.3449913}
    \end{tabular}
    } & \multirow{3}{*}{\begin{tabular}{c}
    \\ Items-Clicks \\ $P(C_i=1 \mid Y, C')$
    \end{tabular}} & 
    \multirow{3}{*}{\begin{tabular}{c} \\ \\ $=$\end{tabular}} &
    \multirow{3}{*}{\begin{tabular}{c} Items up to Chosen Item \\ Clicks above Chosen Item  \\ $ P(C_i=1 \mid Y_{1:i}, C'_{1:i-1})$
    \end{tabular}
    } & 
    \multirow{3}{*}{\begin{tabular}{c} \\ \\ $=$\end{tabular}} & \multirow{3}{*}{\begin{tabular}{c} \\ No Factorization \\ $f\big(Y_{1:i}, C'_{1:i-1}\big)$
    \end{tabular}} \\
    \\ \\ 
     \midrule
    \multirow{3}{*}{
    \begin{tabular}{l}
    \Gls*{drlc} \cite{10.1145/3404835.3463228}
    \end{tabular}
    } & \multirow{3}{*}{\begin{tabular}{c}
    \\  Items-Clicks \\ $P(C_i=1 \mid Y,C')$
    \end{tabular}} & 
    \multirow{3}{*}{\begin{tabular}{c}  \\ \\ $=$\end{tabular}} &
    \multirow{3}{*}{\begin{tabular}{c}   All Items \\ All Clicks \\ $ P(C_i=1 \mid Y,C')$
    \end{tabular}
    } & 
    \multirow{3}{*}{\begin{tabular}{c} \\ \\ $=$\end{tabular}} & \multirow{3}{*}{\begin{tabular}{c}  \\ No Factorization \\ $f(Y,C')$
    \end{tabular}}
    \\ \\  \\
    \midrule
    \multirow{3}{*}{
    \begin{tabular}{l}
    \Gls*{gubm}~\citep{10.1145/3209978.3209990}
    \end{tabular}
    } & \multirow{3}{*}{\begin{tabular}{c}
    \\ Items-Clicks \\ $P(C_{i,j}=1 \mid Y, C')$
    \end{tabular}} & 
    \multirow{3}{*}{\begin{tabular}{c} \\ \\ $=$\end{tabular}} &
    \multirow{3}{*}{\begin{tabular}{c} Chosen Item \\ Clicks with Earlier Timestamps \\ $ P(C_{i,j}=1 \mid Y_{i,j}, \{c \in C' : t(c) < t(C_{i,j})\})$
    \end{tabular}
    } & 
    \multirow{3}{*}{\begin{tabular}{c} \\ \\ $=$\end{tabular}} & \multirow{3}{*}{\begin{tabular}{c} Item Factor \\ Clicks Factor \\ $ f(Y_{i,j}) \cdot g(\{c \in C' : t(c) < t(C_{i,j})\})$
    \end{tabular}} \\
    \\ \\ 
    \midrule
    \multirow{4}{*}{
    \begin{tabular}{l}
    \Gls*{ccm1} \cite{10.1145/3643709}
    \end{tabular}
    } & \multirow{4}{*}{\begin{tabular}{c}
    \\  \\ Fully Dependent \\ $P(C_{i,j}=1 \mid T, Y, C')$
    \end{tabular}} & 
    \multirow{4}{*}{\begin{tabular}{c} \\ \\ \\ $=$\end{tabular}} &
    \multirow{4}{*}{\begin{tabular}{c} Topics up to Chosen Topic \\ Chosen Item \\ Clicks before Chosen Item on the Same Topic \\ $P(C_{i,j}=1 \mid T_{1:i}, Y_{i,j}, C'_{i,1:j-1})$
    \end{tabular}
    } & 
    \multirow{4}{*}{\begin{tabular}{c} \\ \\ \\ $=$\end{tabular}} & \multirow{4}{*}{\begin{tabular}{c} Topics Factor \\ Item Factor \\ Clicks Factor \\ $f(T_{1:i}) \cdot g(Y_{i,j}) \cdot h\big( C'_{i,1:j-1}\big)$
    \end{tabular}} \\
    \\ \\ \\ 
    \bottomrule
  \end{tabular}
\end{adjustbox}
  \vspace{-\baselineskip}
\end{table*}

\section{Designing a Carousel Click Model}

In this section, we illustrate how to design a carousel click model using our three key design choices. %

\heading{Global dependencies}
Following the principle of global dependencies, we plan to develop a model that falls under the category \emph{Topic-Items}, where a click event is influenced by the displayed topics and items, but not by other clicks.
This gives us the general conditional probability:
\begin{equation}
    P(C_{i,j}=1 \mid T, Y, C') = P(C_{i,j}=1 \mid T, Y).
\end{equation}

\heading{Sequentiality}
Furthermore, we assume that a click within the \(i\)-th carousel depends on the preceding and current topics \(T_{1:i}\) and the preceding displayed items for the same topic \(Y_{i,1:j-1}\) and the item itself \(Y_{i,j}\):
\begin{equation}
    P(C_{i,j}=1 \mid T, Y) = P(C_{i,j}=1 \mid T_{1:i}, Y_{i,1:j}).
\end{equation}

\heading{Factorization}
Finally, we apply factorization to decompose the overall click probability into several components.
One possible factorization is given by:
\begin{equation}
    P(C_{i,j}=1 \mid T_{1:i}, Y_{i,1:j}) = f(T_{1:i}) \cdot g(Y_{i,1:j}),
\end{equation}
which separates the effects of topics and displayed items via \(f(\cdot)\) and \(g(\cdot)\), respectively.
This factorization clarifies how the contribution of each component can be decomposed. %

Alternatively, one could also choose a factor \(f'(\cdot)\) for the item itself with a joint function \(g'(\cdot)\) to capture the combined influence of topics and preceding items:
\begin{equation}
    P(C_{i,j}=1 \mid T_{1:i}, Y_{i,1:j}) = f'(Y_{i,j}) \cdot g'(T_{1:i}, Y_{i,1:j-1}).
\end{equation}

\noindent%
By answering the three key design choices, we have developed the most salient mathematical properties of our new carousel click model.
Naturally, there are further choices to be made, e.g., \emph{whether to use \glspl{pgm} or \glspl{nn} for each of the factors}, \emph{what loss function to employ when fitting the model}, etc.
However, with our decisions so far, we have already described what statistical patterns the model can capture and we can thus compare it with existing click models.
Therefore,  we argue our theory-based approach provides a flexible framework that correctly concerns the mathematical capabilities of models first-and-foremost.

\section{Conclusion}
In this paper, we re-established the theoretical foundations and conceptualization of  click models by primarily considering their mathematical properties.
We propose that the categorization of click models should be based on the mathematical relations between observable variables rather than on presumed user behavior or latent variables. 
Building on this proposition, we introduced three key design choices: \emph{global dependencies}, \emph{sequentiality}, and \emph{factorization}. 
We argue that these choices describe the most salient mathematical properties of click models, and therefore, should be the first concern for their categorization, design and comparison.

Our approach results in a taxonomy that covers any - existing or future - single-list, grid or carousel interface click models based on \glspl{pgm}, \glspl{nn} or \glspl{gbdt}.
Uniquely, our structured categorization can meaningfully compare click models of different types and for different interfaces.
Lastly, we apply our theory-based approach to design a novel carousel click model to demonstrate how our design choices provide a framework for the future development of new click models.

Our work provides a new perspective on the theory underlying click model research, and in doing so, reveals that many unexplored opportunities for click models remain, especially for layouts beyond the single list, such as carousel interfaces.

\begin{acks}
This work is supported by the Dutch Research Council (NWO) under grants \href{https://www.nwo.nl/en/projects/viveni222269}{VI.Veni.222.269}, \href{https://www.nwo.nl/en/projects/024004022}{024.004.022}, \href{https://www.nwo.nl/en/projects/nwa138920183}{NWA.1389.20.183}, and \href{https://www.nwo.nl/en/projects/kich3ltp20006}{KICH3.LTP.20.006}.
It is also supported by the European Union's Horizon Europe program under grant agreements No.\ \href{https://doi.org/10.3030/101070212}{101070212} and No.\ \href{https://doi.org/10.3030/101072410}{101072410} (Eyes4ICU, funded under the Marie Sk\l{}odowska-Curie Actions).
All content represents the opinion of the authors, which is not necessarily shared or endorsed by their respective employers and/or sponsors.
\end{acks}

\clearpage

\bibliographystyle{ACM-Reference-Format}
\balance
\bibliography{references.bib}

\end{document}